\documentclass{article}

\begin{document}

\title{On the Casimir Effect in the High Tc Cuprates}
\author{Achim Kempf\\ Departments of Applied Mathematics and Physics \\
 University of Waterloo,
Ontario, Canada}
\date{}
\maketitle
\begin{abstract}
High temperature superconductors have in common that they consist of
parallel planes of copper oxide separated by layers whose composition can
vary. Being ceramics, the cuprate superconductors are poor conductors
above the transition temperature, $T_c$. Below $T_c$, the parallel Cu-O
planes in those materials become superconducting while the layers in
between stay poor conductors. Here, we ask to what extent the Casimir
energy that arises when the parallel Cu-O layers become superconducting
could contribute to the superconducting condensation energy. Our aim here
is merely to obtain an order of magnitude estimate. To this end, the
material is modelled as consisting below $T_c$ of parallel plasma sheets
separated by vacuum and as without a significant Casimir effect above
$T_c$. Due to the close proximity of the Cu-O planes the system is in the
regime where the Casimir effect becomes a van der Waals type effect,
dominated by contributions from TM surface plasmons propagating along the
$ab$ planes. Within this model, the Casimir energy is found to be of the
same order of magnitude as the superconducting condensation energy.

\end{abstract}

\section{Introduction}
The question has remained open how Cooper pairs can be stable at around
100K where some high temperature superconductors (HTSCs) are still
superconducting. In particular, the phonon-mediated attractive
electron-electron interaction of conventional superconductors \cite{bcs}
is too weak at these temperatures. While much is known about the
microscopic mechanism of high temperature superconductivity, for example,
that it involves $d$-wave Cooper pairs \cite{handbook,micro}, the
energetics that drives this mechanism is still unclear
\cite{leggett-energy}. For a new approach to the question of the
energetics behind high temperature superconductivity, let us reconsider a
feature that HTSCs have in common, namely parallel superconducting layers
which are separated by layers of essentially insulating material. Since in
between any two conducting planes there occurs a Casimir effect
\cite{casimir,mostepanenko-book,milton-book}, the effect should also occur
between the parallel superconducting layers in HTSCs, as was first pointed
out in \cite{ak}.

Before estimating the significance of the Casimir effect in HTSCs, let us
recall the textbook case of two parallel plates that are separated by
vacuum and that are ideal conductors, i.e., conductors whose conductivity
and Mei{\ss}ner effect expel electromagnetic fields of all wavelengths
with vanishing penetration depths. If the plates' area, $A$, is large
compared to their distance, $a$, the Casimir energy reads
\begin{equation} E_{c}(a) = -
\frac{\pi^2\hbar cA}{720a^3}, \label{casimir2}
\end{equation}
thus leading to an attractive force. Corrections that take into account
the finite conductivity of real metals have been calculated for geometries
such as parallel plates and a plate and a sphere, along with corrections
for finite surface roughness and finite temperature, see
\cite{casimir-reviews}. Recent experiments measured the Casimir force
between a metallic plate and sphere down to distances of around $100nm$,
confirming the theoretical predictions with a precision of $0.5\%$
\cite{mohideen-exp}.

In the case of HTSCs, as the temperature is lowered below $T_c$,
superconducting charge carriers form in parallel layers. The onset of
superconductivity does not make these Cu-O layers ideal conductors but
even only as superconductors these layers should lead to some extent to a
Casimir effect and therefore to some negative Casimir energy. If this
lowering of the energy at the onset of the superconductivity is large
enough then it could be the very reason why the Cu-O layers' initially
non-superconducting charge carriers are able to form superconducting
charge carriers. Because of the Casimir effect which arises at the onset
of superconductivity, the initially non-superconducting charge carriers
would be energetically driven to use whichever microscopic mechanism is
available to them to form superconducting charge carriers. In this
scenario, it would therefore be the very effects of superconductivity
which enable and stabilize superconductivity. Cooper pairs would derive
their stability collectively, across layers. Namely, Cooper pairs would be
stable because if sufficiently many of the Cooper pairs on opposing layers
were to break up then the Casimir effect would cease and the energy would
have to go back up.

\section{Estimating the size of the Casimir effect in HTSCs}

Our aim now is to estimate if the Casimir effect in HTSCs could indeed be
sufficiently large to make the formation of Cooper pairs energetically
favorable at temperatures as high as 100K. Clearly, the superconducting
Cu-O layers are much less efficient at suppressing electromagnetic fields
than ideal conductors would be, and the more so the shorter the
wavelength. In particular, since the layer spacing is two to three orders
of magnitude smaller than for example the London penetration depth, the
Casimir effect should be suppressed by several orders of magnitude. In
order to estimate the order of magnitude of the actually occurring Casimir
effect in HTSCs, let us crudely model the charge carriers of the
superconducting Cu-O planes in the superconducting state as forming
$2$-dimensional parallel plasma sheets separated by vacuum. For the normal
state we make the approximation that the material does not possess a
significant Casimir effect.

It is known that for two parallel plasma sheets with large separation,
$a$, the Casimir energy as a function of $a$ is given by
Eq.\ref{casimir2}. The Casimir energy for parallel plasma sheets in the
regime of small $a$ has recently been calculated, \cite{bordag}:
\begin{equation}
\tilde{E}_{c}(a) = -5\cdot 10^{-3} \hbar c A
a^{-5/2}\sqrt{\frac{nq^2}{2mc^2\epsilon_0}}\label{bc}
\end{equation}
In our model we are of course in the regime of high layer transparency,
i.e., in the small $a$ regime where Eq.\ref{bc} can be applied. Applying
Eq.\ref{bc} to our situation here, with a typical layer separation of
$a=1nm$, we obtain a Casimir energy which is suppressed by four orders of
magnitude compared to what the Casimir energy would be for two ideally
conducting layers. For example, we obtain
$\tilde{E}_c(1nm)/E_c(1nm)=4.3\cdot10^{-4}$ when using for the charge
carriers' area density, charge and mass the realistic values $n=10^{14}
(cm)^{-2}$, $q=2e$ and $m=2\alpha m_e$ with $\alpha=5$. Let us now
identify $\tilde{E}_{c}(a)$ with the condensation energy
\begin{equation}E_{cond.}=\tilde{E}_c(a)\end{equation}
and calculate the corresponding transition temperature $T_c$. First, we
have
\begin{equation} T_c=\Delta(0)/\eta k_B
\end{equation}
where for HTSCs the parameter $\eta$ is generally thought to be around or
somewhat larger than the BCS value of $\eta=1.76$. Next, we relate the gap
energy to the density of states and to the condensation energy through:
\begin{equation}
E_{cond}= - D(\epsilon_F)\Delta^2(0)/2
\end{equation}
The density of states in the case of a Fermi gas in two dimensions reads
$D(\epsilon_F)=mA/\pi\hbar^2$, so that we finally obtain this prediction
for the transition temperature:
\begin{equation}
T_c(a) = \frac{2^{1/4}\pi^{1/2}\hbar^{3/2}e^{1/2}n^{1/4}}{10\eta k_B
m^{3/4}\epsilon_0^{1/4} a^{5/4}}\label{seven}
\end{equation}
Notice that, curiously, $T_c$ is a function of $n/a^5$. If we choose
realistic values such as $a=1nm, n=10^{14} (cm)^{-2}$, $\eta=1.76$ and
$m=5m_e$, this yields:
\begin{equation}
T_c=125K \label{tem}
\end{equation}
In order to illustrate how strong the Casimir effect would be at such
small layer distances, if the Casimir effect were not suppressed because
of the high layer transparency, let us compare with the result that is
obtained when modelling the Cu-O layers as ideally conducting sheets. In
this case, Eq.\ref{casimir2} would apply instead of Eq.\ref{bc} which
would have led to a prediction of a transition temperature of $T_c=
3350K$. The result for $T_c$ given in Eq.\ref{tem} is of course only a
very rough estimate, but it shows that the Casimir effect could be of the
right order of magnitude. Within our simple model, even though the Casimir
effect is highly suppressed in HTSCs, the small Casimir effect that does
occur could still be large enough to account for the superconducting
condensation energy in HTSCs.

\section{Conclusions}

Within this scenario, the stability of Cooper pairs is a collective
phenomenon as it involves the Casimir effect across layers. More
precisely, it is mostly a van der Waals type effect between the Cu-O
layers: because of the small distance between neighboring layers the
retardation of the electromagnetic interaction which couples the charge
and current fluctuations on one Cu-O layer to the fluctuations on nearby
Cu-O layers are negligible and we are therefore in the regime where the
Casimir effect becomes a van der Waals type effect. Correspondingly, $c$
did in fact drop out of our calculations. Therefore, in this scenario, the
energetic gain at the onset of superconductivity is due to an increase in
the fluctuations along the Cu-O layers that are in sync with fluctuations
on nearby Cu-O layers, which then leads to an increased attractive van der
Waals type force in between the Cu-O planes - with a corresponding drop in
a van der Waals type interaction energy. We can say somewhat more because
it is known that the Casimir effect for parallel plasma sheets is at short
distances dominated by contributions from TM surface plasmons propagating
along the plasma sheets \cite{bordag}. Applied to our scenario, this would
imply that the fluctuations that are most enhanced after the onset of
superconductivity and that therefore contribute most to the energetics of
the Casimir van der Waals effect are the TM ``surface" plasmons
propagating along the superconducting Cu-O layers. These TM surface
plasmons should presumably also play a role in the microscopic mechanism
that allows Cooper pairs to form. We recall that the pairing interaction
is today expected to be of magnetic origin \cite{handbook}.

\section{Outlook}

In order to make testable quantitative predictions, our model needs to be
significantly improved in several respects. First, it will need to be
taken into account that also above $T_c$ the cuprates possess to some
extent a layered pattern of (small) conductivity, probably describable by
a Drude model, leading to a small Casimir energy. The Casimir energy that
is  available as condensation energy is of course only the difference
between the Casimir energies above and below $T_c$.

Also, we considered so far only the case of two layers. Actual HTSCs
possess not only many layers but also multiple layer distances within and
among the units cells. In principle, it should be straightforward to
generalize our present results for multiple layers at multiple distances,
namely by calculating the transmission and reflection coefficients of more
layers from those of fewer layers, iteratively, and if need be
numerically. Given the high transparency of the Cu-O layers also the
coupling between fluctuations of layers whose distance may not be in the
short distance regime may need to be taken into account, in which case
also some Casimir effect of non van der Waals origin would play a role.
These calculations should then offer an opportunity to check our scenario
experimentally.

For example, in $YBa_2Cu_3O_{7-x}$ (YBCO), which becomes superconducting
at around $92K$ the crystallographic unit cell contains two copper oxide
layers at a distance of $a_b \approx 0.39nm$ and neighboring such
bi-layers are separated by a layer of essentially nonconducting material
of width $a_i \approx 1.17nm$. The area density of the superconducting
charge carriers on each Cu-O layer is roughly on the order of $n \approx
10^{17}/m^2$. For our purpose, the case of YBCO is of particular interest
because of the availability of experimental data, \cite{budai,too}, on
epitaxial superlattices in which slabs of YBCO alternate with slabs of
insulating material, namely $PrBa_2Cu_3O_{7-x}$ (PrBCO). For example, in
the experiments reported in \cite{budai}, the authors varied the thickness
of the YBCO slabs from $M=1$ to $M=8$ unit cells and the thickness, $a_m$,
of the PrBCO slabs from $N=1$ to $N=16$ unit cells, i.e., in the range
$a_m=2nm$ to $20nm$. The superconducting transition temperature was
measured as a function, $T_c(N,a_m)$, of $N$ and $a_m$ and it was found
that $T_c$ characteristically increases with decreasing layer distances.
Since our ansatz predicts a dependence of $T_c$ on the layer separation,
Eq.\ref{seven}, those data should provide a good testing ground for the
present ansatz. Given that the Casimir effect diminishes with increasing
layer distances our approach should match the experimental data at least
qualitatively. Work in this direction is in progress.

If our general scenario applies, it would mean that in order to raise
$T_c$ the challenge would be to create layered materials, not necessarily
cuprates, for which the Casimir effect in the superconducting state is as
large as possible. In order to maximize the Casimir energy available as
condensation energy the same material would also have to possess an as
small as possible Casimir effect in the normal state. Of course, for the
Casimir effect to enable superconductivity energetically, the material in
question must first possess some microscopic mechanism that allows
superconducting charge carriers to form in layers. Let us consider, for
example, carbon nanotubes (CN) as these can be made superconducting
\cite{cn1}. Within our scenario, it is to be expected that, due to the
Casimir effect, $T_c$ should be higher for multi-walled than for
single-walled CNs. Indeed, recently, a 30-fold increase in $T_c$ for
multi-walled CNs has been reported \cite{cn2}.

It should also be interesting to explore a possible connection to the
approach in \cite{leggett}. There, the condensation energy has been
suggested to arise from an increased screening of the charge carriers'
Coulomb repulsion at the onset of superconductivity, an effect which is
also characteristically layer-distance dependent.

$$$$\bf Acknowledgement: \rm The author is grateful to Robert Brout,
Michel Gingras and Anthony Leggett for their criticisms. This work has
been supported by CFI, OIT, PREA and the Canada Research Chairs Program of
the National Science and Engineering Research Council of Canada.
$$$$

\vfill
\end{document}